\documentclass[aps,prd,twocolumn,nofootinbib,superscriptaddress%
 reprint,superscriptaddress,
 amsmath,amssymb,longbibliography,aps,prl
]{revtex4-2}

\usepackage{natbib}
\usepackage{graphicx}
\usepackage{dcolumn}
\usepackage{bm}

\usepackage{graphicx}
\usepackage{dcolumn}
\usepackage{bm}
\usepackage{hyperref}
\usepackage{xcolor}
\usepackage{amsmath, amssymb, amsfonts}
\usepackage{mathtools}
\usepackage{subfigure}
\usepackage{mathrsfs}
\usepackage{comment}
\usepackage{amsthm}
\usepackage{ dsfont }

\definecolor{purple1}{rgb}{128,0,128}

\newcommand{\bea}{\begin{eqnarray}}
\newcommand{\ea}{\end{eqnarray}}

\theoremstyle{definition}

\definecolor{dfcol}{cmyk}{1, 0.2108, 0.13, 0.3}
\newcommand{\df}[1]{\ifthenelse{\boolean{}}{\textcolor{dfcol}{[{\bf DF}: #1]}}{}}

\renewcommand{\[}{\begin{equation}}
\renewcommand{\]}{\end{equation}}

\definecolor{mygray}{gray}{0.6}

\begin{document}

\title{Probing Hawking Temperature Threshold via Quantum Depletion in Bose-Einstein Condensate}
\title{Probing Hawking Temperature Threshold via Quantum Depletion in Bose-Einstein Condensate}
\author{Arun Rana}
\affiliation{%
Seoul National University, Department of Physics and Astronomy, Center for Theoretical Physics, Seoul 08826, Korea
}%



\begin{abstract}
We investigate the correlation between quantum depletion and Hawking temperature in a ring-shaped Bose-Einstein condensate featuring an analog black hole–white hole horizon pair, using the Bogoliubov approach. The presence of horizons is found to enhance the quantum depletion compared to horizon-free configurations, indicating a correlation between depletion and horizon dynamics. Via tuning the Hawking temperature, we observe its effect on the depletion profile. Our results show that depletion increases with Hawking temperature, and beyond a certain threshold, backreaction effects emerge, challenging the validity of the Bogoliubov approximation.  We identify a viable parameter regime where the system remains both theoretically controlled and experimentally accessible, offering insight into horizon-induced quantum fluctuations, with implications for future studies of backreaction.
\end{abstract}

\maketitle

\section{Introduction}
Black holes represent some of the most enigmatic objects in the universe, providing an exceptional framework for examining how general relativity and quantum theory intersect. The pioneering research by Bekenstein and Hawking \cite{PhysRevD.7.2333,Hawking:1974rv,Hawking:1975vcx} revealed that black holes demonstrate thermodynamic properties, radiating thermal energy at a specific temperature—what we now call Hawking radiation. Nevertheless, this Hawking temperature remains extraordinarily low for astronomical black holes, creating significant obstacles for direct experimental observation. Additionally, \cite{Hawking:1975vcx} overlooked a crucial element—the backreaction, meaning the influence of quantum fluctuations on the background spacetime. Since the direct investigation of such systems is not possible, these constraints can rather be addressed and investigated in laboratory settings by developing quantum simulators for such phenomena using ultracold atomic gases \cite{Barcelo:2005fc,Schutzhold:2025qna}.
In this context, Bose-Einstein condensates (BECs) emerge as particularly valuable tools, offering precise experimental control alongside well-developed microscopic theoretical frameworks \cite{Cornell:2002zz}. The quantum simulation of such spacetime using superfluids originated from Unruh's proposal in 1981 \cite{Unruh:1980cg}, in which he showed that transonic fluid flows can mimic event horizons and suggested that black hole evaporation processes might be observable in these systems.

While other platforms, such as superfluid helium, nonlinear optics, hydrodynamical analogs, and superconducting circuits \cite{Volovik:2000ua,PhysRevA.86.063821, Gaona-Reyes:2017mks, PhysRevLett.122.010404,Braunstein, Patrick:2018orp,Euve:2015vml,Euve:2018uyo,Shi:2021nkx,Coutant:2016vsf,Svancara:2023yrf,Braunstein:2023jpo,Tian:2018srd,Yang:2024fql}, have been explored, BECs remain particularly compelling \cite{bloch:hal-03740937} and highly favorable platform, with experimental observations of analog Hawking radiation reported in \cite{Steinhauer:2014dra, Steinhauer:2015saa, MunozdeNova:2018fxv,Kolobov2021}, reinforcing the feasibility of using quantum gases as simulators. Numerous theoretical and experimental studies have since examined sonic black holes in BECs \cite{PhysRevLett.105.240401,PhysRevA.63.023611,Barcelo_2006,Macher,PhysRevD.105.124066,PhysRevD.107.L121502,Isoard:2019buh,Kolobov:2019qfs,Fabbri:2020unn,Dudley:2020toe,Balbinot:2021bnp,Jacquet:2021scv,Jacquet:2022vak,Tolosa-Simeon:2022umw,Viermann:2022wgw,Palan:2022sly,Syu:2022cws,Syu:2022ajm,Vieira:2023ylz,Agullo:2024lry,Berti:2024cut,Wang:2016jaj, Michel:2016tog}, as well as early universe analogs \cite{PhysRevA.69.033602, PhysRevA.70.063615, PhysRevLett.91.240407, Rana:2023mgr,Butera:2021rxz,Unruh:2022gso,Butera:2023psb,Eckel,Cha:2016esj,Banik:2021xjn}. Notably, ring-shaped (toroidal) condensates \cite{PhysRevLett.95.143201,PhysRevA.73.041606,PhysRevLett.106.130401,PhysRevA.72.063612,PhysRevA.63.023611,PhysRevA.76.023617} have gained attention for their intrinsic stability and periodic boundary conditions, offering a versatile setup for analog gravity simulations.

Alongside reproducing the kinematic aspects of curved spacetime, BEC-based quantum simulators go beyond and enable the controlled study of microscopic backreaction effects, otherwise inaccessible in real gravitational systems \cite{PhysRevD.72.105005,PhysRevA.106.053319}. Despite the intuitive assumption that Bogoliubov theory could easily handle backreaction in dilute BECs, this turns out to be nontrivial \cite{PhysRevA.106.053319}. Though the broader research goal would be to investigate the correlation between Hawking temperature and backreaction \cite{PhysRevD.72.105005,Balbinot,Butera,PhysRevA.106.053319,Caio_PRA,Pal,Sols} in such systems, via this work, however, we take a crucial fundamental step by examining the correlation between Hawking temperature $T_H$ and quantum depletion—the fraction of atoms that exit the condensate due to interactions and dynamic evolution.

\begin{figure}[h]
\centering
\includegraphics[width=0.4\textwidth]{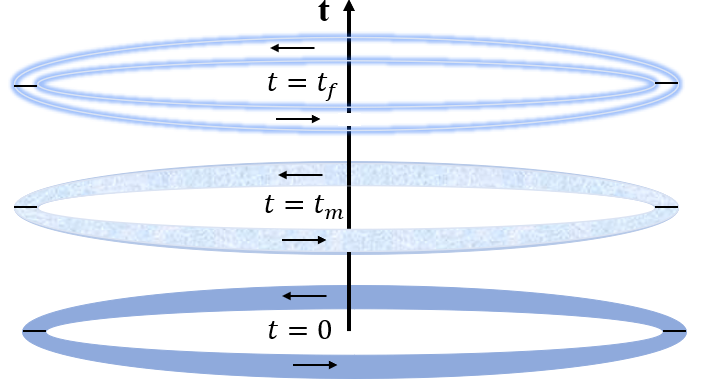}
\caption{Artist's impression of ring BEC featuring acoustic black hole (left) and white hole (right) horizons. The state at $t = 0$ corresponds to an initial configuration with no quantum depletion. As time progresses to $t = t_m$, quantum depletion begins to evolve, eventually reaching a maximal value at $t = t_f$, beyond which the background can no longer be considered a condensate, with the quantum component dominating over the coherent background that characterized the initial state.}
\label{fig:cartoon_sketch}
\end{figure}

In this paper, we study a one-dimensional ring-shaped Bose-Einstein condensate hosting an analog black hole (BH) and white hole (WH) horizon pair \cite{Yatsuta}, as depicted in Fig.~\ref{fig:cartoon_sketch}. This geometry provides a controlled and stable background for simulating Hawking-like processes, allowing us to dynamically evolve the system using the time-dependent Gross-Pitaevskii equation (GPE) and analyze quantum fluctuations via the Bogoliubov–de Gennes (BdG) formalism. The initial state is a fully condensed vacuum with no depletion, evolving in time as interactions and Hawking radiation-induced excitations generate depletion.

Our primary aim is to determine the threshold of $T_H$ beyond which the condensate ceases to be well-described by Bogoliubov theory—a limit corresponding to significant quantum depletion and the onset of strong backreaction. We find that quantum depletion increases monotonically with $T_H$, and that once a critical level is reached, the coherence of the condensate is lost, giving way to a regime dominated by quantum fluctuations. The structure of the article is as follows: in Section~\ref{sec2}, we introduce the theoretical framework and construct the analog model consisting of a blackhole-whitehole pair in a ring condensate with horizon and background dynamics being defined in Section~\ref{sec2a}, alongside Section~\ref{sec2b} being dedicated to the study of quantum depletion dynamics. In Section~\ref{sec3}, we analyze the analog Hawking temperature and its influence on the quantum depletion. Finally, we present our conclusions and future directions in Section~\ref{sec4}, ending the article in Section~\ref{sec5} with the acknowledgement.





\section{Construction of black hole-white hole pair on a ring BEC}  \label{sec2}
 We consider a non-relativistic dilute bosonic gas, described by the field $\phi =\phi(t,x)$ in $1+1D$, which on the mean-field level is described by the GPE 
 $(\hbar = 1)$
\begin{equation} \label{eq:gpe}
    i\partial_t\phi=-\frac{1}{2m}\partial^2_x\phi+(V+g|\phi|^2)\phi.
\end{equation}
with $m$ the boson mass, $g$ the coupling strength, and 
$V$ representing the external potential. We prepare the condensate to be in a ring-like configuration of total length $l$, satisfying periodic boundary conditions $\phi(t,-l/2)=\phi(t,l/2)$ and $\partial_x\phi(t,x)|_{x=-l/2}=\partial_x\phi(t,x)|_{x=l/2}$. We assume that for $t<0$, the condensate density, $\rho_0$ is constant, $g=0$, and then switch to a $g_0>0$.  
Accordingly, using (\ref{eq:gpe}), 
\begin{equation}
    \frac{1}{g_0\rho_0}i\partial_t\phi=-\frac{\xi_0^2}{2}\partial_x^2
\phi+\bigg(\frac{V}{g_0\rho_0}+\frac{g|\phi|^2}{g_0\rho_0}\bigg)\phi,
\end{equation}
where we defined $\xi_0=1/\sqrt{mg_0\rho_0}$. In what follows, we use as the unit of length $\xi_0$, thus time in units of 
$\xi_0^2$, 
$g$ is in units $g_0$ and $\phi$ in units of $\sqrt{\rho_0}$. 
Then we have 
\begin{equation}
    i\partial_t\phi=-\frac{1}{2}\partial^2_x\phi+(V+g|\phi|^2)\phi,
\end{equation}
where normalization is imposed, 
  $  \int^{l/2}_{-l/2}dx|\phi|^2=N$ with as the number of particles in the condensate.
For $t<0$, the condensate is given by
   $ \phi(t,x)=e^{-i\mu t+ik_nx}$ 
with the chemical potential $\mu$. Boundary conditions at $x=\pm l/2$ lead to 
   $ k_n=\frac{2n\pi}{l}$, 
where $n=0,1,2,....$, whereas the GPE implies 
   $ \mu =\frac{k_n^2}{2}+V.$ 
This fixes our external potential. The interaction $g(x)$  we take to be Gaussian,  
\begin{equation}
    g(x)=1-\frac{e^{-x^2}}{2}. \label{Gaussiang(x)}
\end{equation}
 and is turned on at $t>0$, which completes our setup for constructing the black hole-white hole (BH-WH) pair in our ring condensate.


\subsection{Horizons and background evolution} \label{sec2a}
\begin{figure}[t]
    \includegraphics[width=0.44\textwidth]{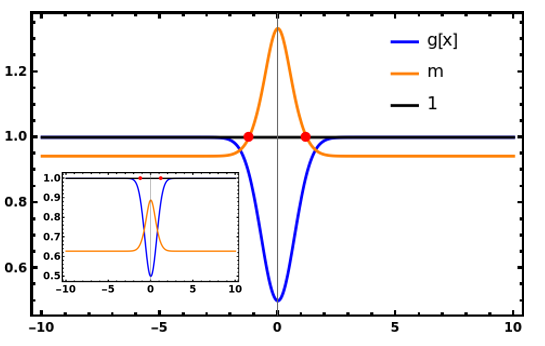}
    \caption{mach number (orange)
    and g(x) (blue) vs x for n=3. The black line represents $y=1$ and is used for reference. The red dot represents the points on the x-axis where the Mach number crosses one, hence implying the presence of acoustic black hole and white hole horizons.}
    \label{fig:machno_horizons_no_horizons_combined}
\end{figure}

To isolate the effect of the horizons on quantum depletion, we explore 
our basic setup's dynamical response in two cases: The BH-WH pair is present or it is not.  
Having fixed $V$ and $g(x)$, the local Mach number ${\mathfrak M}=v/c$, with $v$ and $c$ being flow velocity and sound speed, respectively, is  
\begin{equation} \label{eq:machno}
  {\mathfrak M}  =\frac{k_n}{\sqrt{g|\phi|^2}}
\end{equation}
at $t=0$, with ${\mathfrak M}=1$ representing the location of analogue event horizons. The presence or absence of the acoustic horizons can be controlled via the parameter $n$ in  $k_n=\frac{2n\pi}{l}$, 
considering that the background density i.e. $|\phi|^2$ and $l$ remains fixed. The length of the ring $(l)$ is taken to be 20. Mach number and interaction $g(x)$ are shown in Figs.\ref{fig:machno_horizons_no_horizons_combined}. When we consider $n=3$, it establishes the BH-WH horizons at the red dot positions that represent the points on the ring where the Mach number reaches unity coming from below (BH) and above (WH).  
Horizons are absent when substituting $n=2$, cf.~inset of Fig.~\ref{fig:machno_horizons_no_horizons_combined}.


\begin{figure}[t]
    \includegraphics[width=0.44\textwidth]{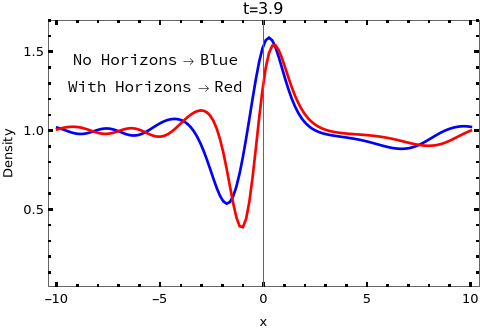}
    \caption{Background condensate for $n=2$ (blue) when there are no horizons, and for $n=3$ (Red) when horizons are present.}
    \label{fig:final_background}
\end{figure}

We now proceed to investigate the dynamics
 of the condensate  for both scenarios,  then adding quantum depletion dynamics later on. 
 For $t>0$, we take the ansatz 
   $ \phi(t,x)=e^{-i\mu t+ik_nx} \varphi$, 
where $\varphi$ is a solution of 
\begin{equation} \label{eq:background}
    i\partial_t\varphi=-\frac{1}{2}\partial^2_x\varphi-ik_n\partial_x\varphi+g|\phi|^2\phi,
\end{equation}
subject to the initial condition $\varphi(0,x)=1$.To study the evolution of background, we solve the equation \ref{eq:background} numerically 
when the BH-WH pair is present and when it is not. We show the results obtained for $|\phi|^2$ on solving (\ref{eq:background}), in Fig.~\ref{fig:final_background}
We conclude from Fig.~\ref{fig:final_background} that $|\phi|^2$ displays a similar behavior for both cases.
Here, we ran the simulation until $t=3.9$; our background is rendered stable in the time window $[0,3.9]$; 
$\varphi(0,x)=1$ was taken to be the initial solution.

\subsection{Quantum depletion} \label{sec2b}
\begin{figure}[h!]
    \centering
    \includegraphics[width=0.44\textwidth]{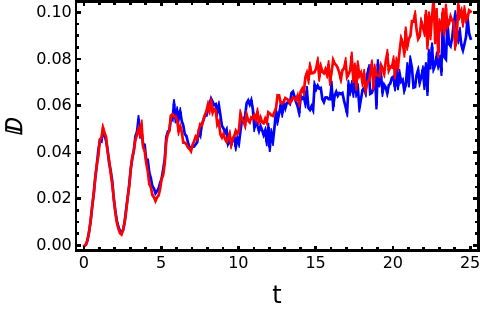}
   \caption{Global quantum depletion $(\mathfrak D)$ as a function of time with (red) and without (blue) horizons.}
    \label{fig:combined_global_depletion}
\end{figure}

Introducing a small fluctuating field $\chi$ on top of the background condensate $\phi$ as $\phi\rightarrow\phi+\chi$, using \ref{eq:gpe}, 
we obtain the following Bogoliubov-de Gennes (BdG) equation for $\chi$
\begin{equation} \label{eq:BdG_eq_chi}
    i\partial_t\chi=-\frac{1}{2}\partial^2_x\chi+(V+2g|\phi|^2)\chi+g\phi^2\chi^*.
\end{equation}
Here, we let $\chi=exp(-i\mu t+ik_nx)\psi$, such that 
\begin{equation}
     i\partial_t\psi=-\frac{1}{2}\partial^2_x\psi-ik_n\partial_x\psi+2g|\varphi|^2\psi+g\varphi^2\psi^*.
\end{equation}
We define the density of the depleted cloud as $\rho_\chi=\langle\hat{\chi}^\dagger\hat{\chi}\rangle$.
Using the Nambu spinor $\Psi$, defined as $\Psi = \begin{pmatrix} \psi \\ \psi^* \end{pmatrix}$, such that (\ref{eq:BdG_eq_chi}) can be written as
\begin{equation} \label{spinor_eq}
    i\partial_t\Psi=-\frac{1}{2}\partial^2_x\Psi-ik_n\sigma_3\partial_x\Psi+2g|\varphi|^2\Psi+g\sigma_\varphi\Psi^*,
\end{equation}
where 
$
\sigma_\varphi = \begin{pmatrix} 
0 & \varphi^2 \\ 
\varphi^{*2} & 0 
\end{pmatrix}.
$ 
For $t<0$, such a set of modes is 
$
\Psi = \frac{1}{\sqrt{l}}e^{-i\omega_jt+ik_jx} \begin{pmatrix} 1 \\ 0 \end{pmatrix}
$
with $j$ any integer, 
and the frequencies 
$
    \omega_j=\frac{k_j^2}{2}+k_jk_n.
$
Here, $\Psi_j$ will evolve as the interactions are turned on ($t>0$). To calculate $\rho_\chi$, we proceed to solve 
Eq.~\ref{spinor_eq} numerically with the initial condition 
$
\Psi(0,x) = \frac{1}{\sqrt{l}}e^{ik_jx} \begin{pmatrix} 1 \\ 0 \end{pmatrix}
$
We then define the {\em global} quantum depletion $\mathfrak D$ as
\begin{equation}
   {\mathfrak D}=\frac{1}{l}\int^{l/2}_{-l/2}\rho_\chi dx
\end{equation}
and calculate it by evaluating $\rho_\chi$ via the solution of (\ref{spinor_eq}), and hence of (\ref{eq:BdG_eq_chi}).
The results obtained are shown in Fig.~\ref{fig:combined_global_depletion}. Again, the red curve represents the evolution of depletion when a BH--WH pair is present in the condensate, while the blue curve corresponds to the case without horizons. This comparison clearly indicates that the presence of acoustic horizons enhances depletion: we obtain a more depleted cloud when horizons are present. As can be seen in Fig.\ref{fig:combined_global_depletion}, the early-time behavior i.e. the initial evolution of the depletion is same for both the cases. This is due to the mean-field interaction effects dominating the stresses caused by Hawking emission from the horizons and can be well understood from our formalism. In the BdG equations, the depletion dynamics arise from two sources: (i) the mean-field interaction term proportional to the coupling strength $g$, and (ii) quantum fluctuations associated with the acoustic horizons. At early times, both systems (with and without horizons) experience similar interaction-driven dynamics because the Hawking emission---which originates from quantum fluctuations near the horizons---requires time to accumulate and propagate throughout the system before significantly affecting the global depletion. The characteristic timescale for Hawking-emitted phonons to influence the global depletion scales as $L/c_s$, where $L$ is the system size and $c_s$ is the sound speed. In contrast, the mean-field interaction-driven dynamics are established throughout the condensate starting from the initial conditions. Only after Hawking-emitted phonons have propagated across the system and accumulated, does the distinction between the two cases become appreciable. This distinction is controlled by flow velocity and condensate density through eqns.\ref{eq:background}-\ref{spinor_eq} and analog Hawking temperature (discussed in next section). Though the distinction between the horizonfull and horizonless cases in Fig.~\ref{fig:combined_global_depletion} may appear subtle due to the oscillatory nature of the solutions, we define and analyze a more weighted quantity `the depletion rate' in Sec.~\ref{sec3} and properly assess the horizon-induced enhancement examining its parametric dependence and correlation with Hawking temperature, demonstrating clear signatures of the acoustic horizon effects in Fig.\ref{fig:final_depletion_vs_th_vs_t} and Fig.\ref{fig:rate_vs_th}.
The oscillations observed in Fig,\ref{fig:combined_global_depletion} for both cases are physical and reflect the quasi-periodic exchange of atoms between the condensate and non-condensate fractions.

\begin{figure}[h!]
    \centering
    \includegraphics[width=0.42\textwidth]{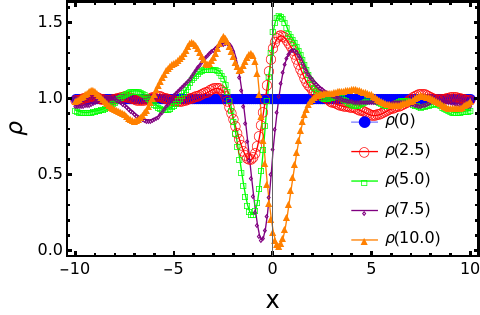}
   \caption{Evolution of the condensate density at different times for $T_H = 0.027mc_0^2$.}
    \label{fig:background_evo}
\end{figure}

\section{Hawking temperature limit and Depletion Rate} \label{sec3}
The previous section demonstrated that the presence of the horizons impacts the depletion magnitude. We now relate the latter to the Hawking temperature. 
We also  aim in this connection to determine the intimately related conditions under which Bogoliubov-de Gennes (BdG) theory remains valid. 
The analog Hawking temperature $(T_H)$,  
which in a quasistationary limit is obtained in the 
horizonful configuration to which we quench \cite{VisserCQG,Carusotto:2008ep}
$(k_B=1)$
\begin{equation}
    T_H=-\frac{\hbar}{2\pi}\frac{1}{2v_h}\left.\frac{d}{dx}(c^2-v^2)\right|_{x=x_H}
\end{equation} \label{eq:hawking_temp}
For our paramters $\rho_0=1$, $l=20$, we find $T_H\sim 0.027$ in units of $mc_0^2=1/\xi_0^2$.
For $T_H \sim 0.027mc_0^2$, the quantum global depletion evolution is shown in Fig.~\ref{fig:combined_global_depletion}. It reaches to value of $0.10$ for this $T_H$. The background evolution until $t=10$ is shown in Fig.~\ref {fig:background_evo}.
This temperature is significantly 
lower than experimentally reported ones \cite{Steinhauer:2014dra, Steinhauer:2015saa, MunozdeNova:2018fxv,Kolobov2021}.

\begin{figure}[h]
    \centering
    \begin{minipage}{0.44\textwidth}
        \centering
        \includegraphics[width=\textwidth]{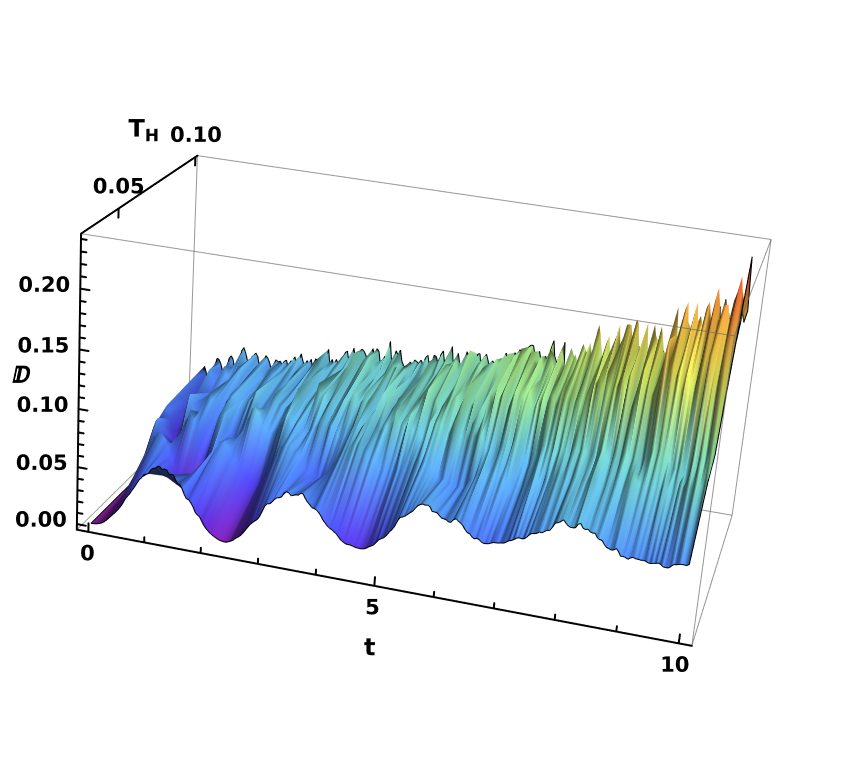}
    \end{minipage}%
    \hspace{-1.5em}
    \begin{minipage}{0.04\textwidth}
        \centering
        \includegraphics[width=1.1\textwidth]{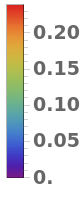}
    \end{minipage}
   \caption{Time evolution of global quantum depletion $(\mathfrak D)$ for various values of $T_H$. The color bar on the right represents the corresponding depletion scale.}
    \label{fig:final_depletion_vs_th_vs_t}
\end{figure}

To explore the effect of the Hawking temperature, we gradually increase $T_H$ by varying the background density $\rho_0$ by up to a factor of ten, while maintaining the acoustic black–white hole horizon pair within the ring condensate. The resulting global depletion is shown in Fig.~\ref{fig:final_depletion_vs_th_vs_t} as a function of both $T_H$ and the time $t$ after quenching to the BH–WH configuration. As expected in the presence of horizons, the global depletion increases monotonically with time, and higher values of $T_H$ lead to a faster accumulation of depleted atoms, producing larger total depletion at a given observation time. Since the absolute value of the global depletion depends on the duration of the evolution, this figure serves as an illustrative visualization of the cumulative effect.  


To quantitatively characterize the depletion dynamics and disentangle transient quench effects from the steady Hawking-induced particle production, we model the time evolution of the global depletion using the fitting form
\begin{equation}
N_{\mathrm{dep}}(t)
= a + \Gamma\, t + A\, e^{-\gamma t}\cos(\omega t + \phi).
\label{eq:depletion_fit}
\end{equation}
This expression consists of three physically distinct contributions. The constant offset $a$ accounts for the initial depletion present immediately after the quench to the BH--WH configuration. The linear term $\Gamma\, t$ represents the secular growth of depletion due to continuous Hawking emission and defines the asymptotic depletion rate. The remaining term describes a damped oscillatory transient arising from quench-induced excitations and finite-size mode interference, with amplitude $A$, damping rate $\gamma$, oscillation frequency $\omega$, and phase $\phi$.

At long times, $t \gg \gamma^{-1}$, the oscillatory contribution decays exponentially and the depletion dynamics are dominated by the linear term, $N_{\mathrm{dep}}(t) \simeq a + \Gamma t$. This allows the depletion rate $\Gamma = dN_{\mathrm{dep}}/dt$ to be extracted in a well-defined and model-independent manner, independently of short-time transient behavior. In contrast, the total depletion necessarily depends on the time interval considered and is therefore not an intrinsic measure of the emission process.

The fitting parameters are initialized directly from the numerical data: the offset $a$ is estimated from the initial depletion, $\Gamma$ from the slope of the late-time evolution, $A$ from the early-time oscillation amplitude, $\gamma$ from the decay timescale of the transient envelope, and $\omega$ from the dominant oscillation period. This procedure ensures stability of the nonlinear fits and minimizes sensitivity to initial guesses.

\begin{figure}[h!]
    \centering
    \includegraphics[width=0.44\textwidth]{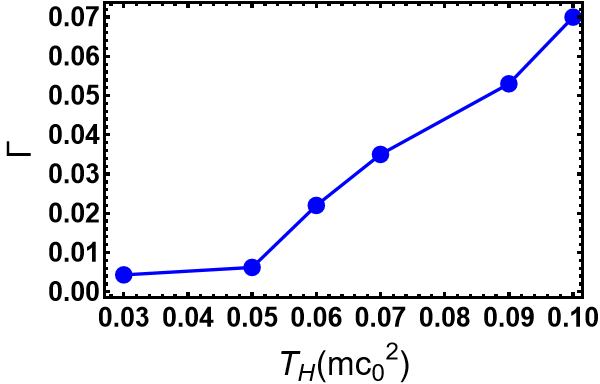}
   \caption{Depletion rate $(\Gamma)$ as a function of Hawking temperature.}
    \label{fig:rate_vs_th}
\end{figure}

For all values of the Hawking temperature considered, the model in Eq.~(\ref{eq:depletion_fit}) reproduces the numerical data with an excellent coefficient of determination, $R^2 > 0.995$. This confirms that the depletion dynamics are well captured by a secular linear growth superimposed with a damped transient and validates the use of $\Gamma$ as a robust and physically meaningful observable. The extracted depletion rates are then used to systematically analyze the dependence of quantum depletion on the Hawking temperature which leads to Fig.~\ref{fig:rate_vs_th}, providing a clear and quantitative characterization of the Hawking-induced depletion across the range of temperatures considered.

On increasing the $T_H$ up to $0.1$ in units of $mc_0^2$, the depletion reaches approximately $25\%$, cf.~Fig.~\ref{fig:final_depletion_vs_th_vs_t}. This choice of threshold is based on the standard assumption that the Bogoliubov-de Gennes (BdG) framework is valid when the fraction of non-condensed atoms is small, $N_{\rm dep}/N \ll 1$ \cite{Pethick_Smith_2008}. Once the depletion reaches a few tens of percent, linearization around the condensate wavefunction becomes increasingly inaccurate, as higher-order correlations and backreaction effects begin to play a significant role. At the 25\% depletion level, around $75\%$ of the particles still occupy the condensate mode, and the system can still be considered predominantly condensed, but beyond-mean-field effects are no longer negligible. Any further increase in $T_H$ accelerates the growth of depletion, eventually rendering the BdG approximation invalid and necessitating a fully quantum many-body or stochastic field-theoretic treatment. The accelerated depletion also enhances backreaction effects, with quantum excitations beginning to influence the background dynamics non-negligibly. A notable advantage of observing depletion at higher values of $T_H$, still below the upper threshold of this study ($T_H \sim 0.1\, mc_0^2$), is that experimental detection of quantum depletion becomes more feasible. The depletion increases rapidly, making it easier to distinguish it from background noise or from finite-temperature effects within the experimental runtime. We also explored an alternative method to increase $T_H$ by spatial variation of the interaction strength $g(x)$ in \eqref{Gaussiang(x)}. 
This has led, however, to the horizons moving too close to one another, thus creating a 
``quasi-microscopic" BH-WH pair  and
making it difficult to resolve Hawking-like effects or local features of the quantum depletion. Modulating the initial condensate density proved to be a more effective and controllable strategy for increasing $T_H$ 
while preserving the global structure around the ring, allowing for a clearer observation of the depletion dynamics with time.

\section{Conclusion and outlook} \label{sec4}
In this work, we have investigated the correspondence between the Hawking temperature in a ring-shaped Bose–Einstein condensate (BEC) hosting an analogue black hole–white hole (BH–WH) pair and the resulting quantum depletion during the system's dynamical evolution within the Bogoliubov framework. By explicitly comparing depletion dynamics in configurations with and without horizons, we demonstrated that the presence of the BH–WH pair leads to a systematically enhanced growth of depletion, revealing a clear connection between horizon-induced phonon production and the loss of condensate coherence.

Our analysis reveals distinct temporal regimes in the depletion dynamics. At early times, the evolution is dominated by mean-field interaction effects proportional to the coupling strength $g$, which act instantaneously throughout the condensate. Both horizonful and horizonless configurations exhibit similar behavior during this phase. The characteristic signature of Hawking emission emerges only after a timescale of order $L/c_s$ (where $L$ being the system size and $c_s$ as the sound speed), the time required for Hawking-emitted phonons to propagate across the system and affect global depletion. This temporal separation between interaction-driven and horizon-induced dynamics provides a clear physical understanding of the observed evolution patterns within our formalism.

A central objective of this study was to identify the physically relevant observable for characterizing Hawking-induced depletion and to explore its dependence on the Hawking temperature $T_H$ within the validity regime of the Bogoliubov–de Gennes (BdG) description. Following a systematic analysis of the long-time evolution, we find that the depletion rate $\Gamma$—rather than the cumulative global depletion—is the appropriate quantity for this purpose. The global depletion grows monotonically in the presence of horizons and necessarily depends on the evaluation time window while the depletion rate, extracted from the secular linear growth in the stationary regime, provides a robust and physically meaningful characterization. The long-time data are well described by linear growth with superimposed damped oscillations, and we identify $\Gamma$ as the coefficient of the linear term. Our fits exhibit excellent quality ($R^2 \simeq 0.995$ for all cases), and the extracted $\Gamma$ increases monotonically with $T_H$, demonstrating unambiguously that the Hawking temperature controls the emission rate. The oscillatory behavior observed in the depletion dynamics reflects the quasi-periodic exchange between condensate and non-condensate fractions and persists even under grid refinement, confirming its physical origin.

Using the background density as a controlled tuning parameter, we explored a broad range of $T_H$ values and established that a quantitative and reliable analysis of depletion dynamics remains possible up to $T_H \sim 0.1\, mc_0^2$. At this temperature, the depletion reaches approximately $25\%$, implying that roughly $75\%$ of the particles still occupy the condensate mode. The choice of this threshold is consistent with the standard criterion that the BdG framework is valid when the fraction of non-condensed atoms is small \cite{Pethick_Smith_2008}. At the $25\%$ depletion level, the system remains predominantly condensed, but beyond-mean-field effects and higher-order correlations are no longer negligible, which justifies the transition to a more complete many-body description. This benchmark is consistent with previous studies of finite-temperature and interaction-induced depletion in BECs and provides a practical criterion for the validity of the BdG approximation in finite-depletion regimes. Further increases in $T_H$ lead to an accelerated growth of depletion, increasingly rapid background evolution, and the onset of significant backreaction effects, which ultimately challenge the applicability of the BdG framework. Once the depletion becomes comparable to the condensate density, the Gross–Pitaevskii plus Bogoliubov description breaks down, and a fully many-body treatment or stochastic field-theoretic approach becomes necessary.

We have also examined different strategies for controlling the Hawking temperature. The parametric dependence of the distinction between horizonful and horizonless configurations is controlled by the flow velocity and condensate density through the horizon formation conditions and the analogue Hawking temperature relation. While modulation of the interaction strength $g(x)$ can, in principle, increase $T_H$, we find that it induces undesirable shifts in the horizon positions, compressing the BH–WH region and limiting spatial and temporal resolution. In contrast, tuning the background density provides a more controlled and robust method for increasing $T_H$ while preserving the global horizon structure, making it better suited for both theoretical modeling and experimental implementations aimed at probing quantum depletion dynamics.

Overall, our results highlight an intrinsic trade-off: higher Hawking temperatures enable faster and more accessible observation of depletion effects, potentially within experimentally realistic timescales, but simultaneously push the system toward regimes where strong backreaction and the breakdown of condensation become unavoidable. By moving towards a systematic analysis of depletion rates and their parametric dependence on $T_H$, this work establishes a general framework for understanding horizon-induced quantum phenomena in simulating gravity systems. This represents a step toward realizing a tabletop analogue-gravity experiment capable of probing aspects of Hawking radiation that are otherwise inaccessible in direct observations \cite{Hawking:1974rv,Hawking:1975vcx}. Future investigations will focus on extending the analysis beyond the Bogoliubov approximation, explicitly incorporating strong quantum backreaction effects, and exploring the microscopic structure and evolution of depletion in the vicinity of the BH–WH pair to gain deeper insight into horizon-induced quantum phenomena.


\section{ACKNOWLEDGMENTS} \label{sec5}
AR gratefully acknowledges Prof. Uwe R. Fischer for conceiving the original idea for this project and for his continuous guidance and insightful contributions throughout all stages of this work, including detailed feedback on the manuscript. Special thanks are extended to Dr. Caio C. Holanda Ribeiro for his input on the development of the initial setup and for many valuable discussions that significantly shaped the progress of the project. This work was supported by the National Research Foundation of Korea under Grant No.~2020R1A2C2008103.


\bibliography{qdH_v1}

\end{document}